\documentclass[pra,twocolumn,showpacs]{revtex4}
\usepackage{graphicx}
\usepackage[usenames]{color}

\newcommand{\beq}{\begin{equation}}
\newcommand{\eeq}{\end{equation}} 
\newcommand{\beqa}{\begin{eqnarray}}
\newcommand{\eeqa}{\end{eqnarray}}
\newcommand{\ba}{\begin{array}}
\newcommand{\ea}{\end{array}}

\begin{document}

\title{Low-temperature thermodynamics of the unitary Fermi gas: \\
superfluid fraction, first sound and second sound}
\author{Luca Salasnich} 
\affiliation{INO-CNR and Dipartimento di Fisica ``Galileo Galilei'', 
Universit\`a di Padova, Via Marzolo 8, 35122 Padova, Italy} 

\begin{abstract} 
We investigate the low-temperature thermodynamics of the 
unitary Fermi gas by introducing a model 
based on the zero-temperature spectra of both bosonic collective 
modes and fermonic single-particle excitations. 
We calculate the Helmholtz free energy and from it 
we obtain the entropy, the internal energy and the chemical potential 
as a function of the temperature. 
By using these quantities and the Landau's expression for the superfluid 
density we determine analytically the superfluid fraction, 
the critical temperature, 
the first sound velocity and the second sound velocity. 
We compare our analytical results with other theoretical predictions 
and experimental data of ultracold atoms and dilute neutron matter. 
\end{abstract} 


\maketitle

\section{Introduction}

In a system of fermions the unitary regime is the situation 
in which $r_e \ll n^{-1/3} \ll |a|$, where $n$ is total number density, 
$r_e$ is the effective radius of the interaction potential 
and $a$ is the s-wave scattering length \cite{giorgini,levin}. 
Thus the system is dilute but the s-wave scattering 
length $a$ greatly exceeds the average interparticle 
separation $n^{-1/3}$. 
It was shown experimentally with dilute and ultracold atomic 
vapors that such systems exist and are 
(meta)stable \cite{hara}. 
It has been suggested that also the dilute neutron matter, 
which is predicted to fill the crust of neutron stars \cite{star},  
is close to the unitary Fermi gas at a certain density range 
\cite{pethick}. 
At low temperature, the thermodynamic properties of 
the superfluid unitary Fermi gas can be obtained from the 
spectrum of elementary excitations, as done many years ago by 
Landau with the superfluid $^4$He \cite{landau0,landau,khalatnikov,abrikosov}. 
This approach has been adopted by Bulgac, Drut and Magierski 
\cite{thermo-bulgac} and also by Nishida \cite{nishida} 
to calculate the internal energy and the entropy of the unitary Fermi gas. 
It has been also suggested by Haussmann, Punk and Zwerger \cite{hpz},  
who proposed a way to calculate the lifetime of fermionic 
excitations at zero temperature. 

In this paper we adopt the Landau 
approach \cite{landau0,landau,khalatnikov,abrikosov} by introducing 
a thermodynamical model which uses the collective bosonic excitations of 
the generalized hydrodynamics \cite{etf} and the spectrum of fermionic 
single-particle excitations \cite{magierski1,magierski2}. 
We calculate the Helmholtz free energy of the two-component balanced 
unitary Fermi gas and from it we determine the entropy, 
the internal energy and the chemical potential. 
In addition, we use the Laudau's criterion to derive the superfluid 
fraction and estimate the critical temperature of the system. 
Finally, by using the obtained superfluid fraction and 
equations of state we calculate 
the first sound and the second sound of the unitary gas 
as a function of the temperature. 
Our results are compared with previous theoretical 
predictions \cite{thermo-bulgac,hu,troyer,taylor,polish,trivedi,wett} 
and experimental data \cite{exp1,exp2,exp-new,jap,nascimbene}. 

\section{Collective and single-particle excitations} 

For any many-body system the weakly excited 
states, the so-called elementary excitations, 
can be treated as a non-interacting gas of 
excitations \cite{landau,abrikosov}. 
In general, these elementary excitations are the result of collective 
interactions of the particles of the system, and therefore pertain 
to the system as whole and not to its 
separate particles \cite{landau,abrikosov}. 
For the unitary Fermi gas the mean-field extended BCS theory 
predicts the existence of fermionic 
single-particle elementary excitations characterized by 
an energy gap $\Delta$ \cite{giorgini,levin}. 
The inclusion of beyond-mean-field 
effects, namely quantum fluctuations of the order parameter, 
gives rise to bosonic collective excitations \cite{giorgini,levin}, 
which are density waves reducing to 
the Bogoliubov-Goldstone-Anderson mode in the 
limit of small momenta \cite{etf}. 

The detailed properties of these elementary excitations 
strongly depend on the approximations involved in the 
theoretical approach \cite{giorgini,levin}. 
As previously stressed, in this paper we extract the details of the 
zero-temperature elementary excitations from a 
density functional approach based on Fixed-Node Diffusion 
Monte Carlo calculation \cite{etf} 
and from recent Path Integral Monte Carlo simulations 
\cite{magierski1,magierski2}. 

It is now well-established \cite{giorgini,levin,nishida} that 
the ground-state energy $E_0$ of the 
uniform unitary Fermi gas made of $N$ atoms in a volume $V$ is given by 
\beq 
E_0  = {3\over 5} \xi  N  \epsilon_F  
\label{energy-0}
\eeq 
where $E_0$ is the ground-state internal energy, 
$\xi\simeq 0.4$ is a universal parameter \cite{bertsch} and 
$\epsilon_F= \hbar^2(3\pi^2 n)^{2/3}/(2m)$ is the Fermi energy 
with $n=N/V$ the number density and $N$ the number of atoms 
of the uniform system in a volume $V$. 

\begin{figure}
\includegraphics[width=1.\linewidth,clip]{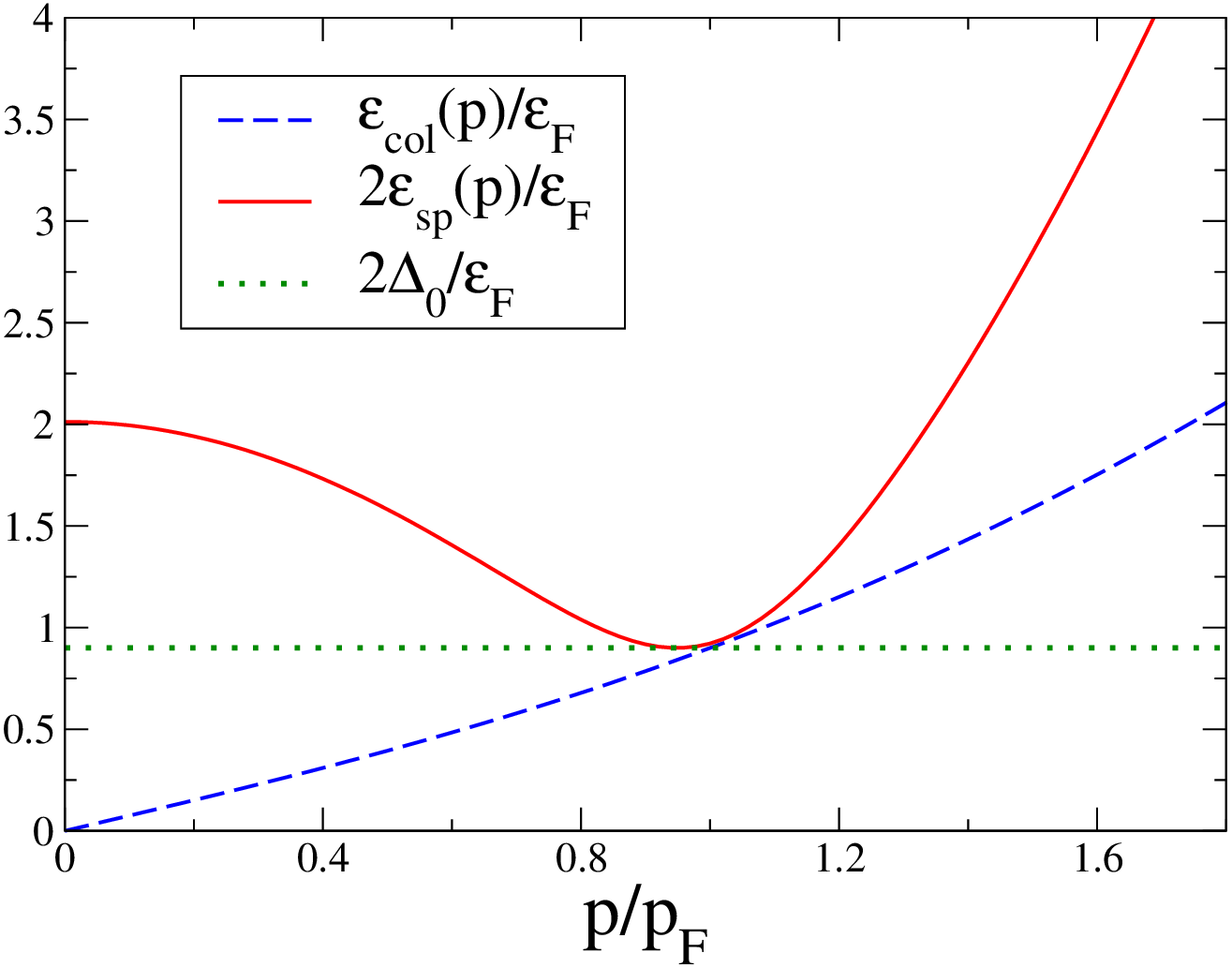}
\caption{(Color online). 
Elementary excitations of the unitary Fermi gas: 
bosonic collective excitations $\epsilon_{col}(p)$ 
(dashed line) and fermionic single-particle excitations 
$2\epsilon_{sp}(p)$ (solid line). 
The collective mode $\epsilon_{col}(p)$ decays in the single-particle 
continuum when there is the breaking of Cooper pairs, 
namely above $\epsilon_{th}=2\Delta_0$ (dotted line). 
Zero-temperature parameters of elementary 
excitations: $\xi=0.42$, $\lambda=0.25$, $\zeta=0.9$, and $\gamma=0.45$.} 
\label{f1}
\end{figure}

The exact dispersion relation of elementary 
(collective and single-particle) excitations is not fully known 
\cite{giorgini,levin}. In Ref. \cite{etf} we have found 
the dispersion relation of collective elementary excitations as 
\beq
\epsilon_{col}(q) = \sqrt{c_1^2 q^2 + {\lambda\over 4m^2}q^4 } \; , 
\label{disp-col}
\eeq
where 
\beq 
c_1 = \sqrt{\xi\over 3}\ v_F  \; , 
\label{disp}
\eeq
is the zero-temperature first sound velocity, 
with $v_F=(\hbar/m)(3\pi^2 n)^{1/3}$ the 
Fermi velocity of a noninteracting Fermi gas (see Fig. \ref{f1}). Notice that 
the term with $\lambda$ takes into account the increase 
of kinetic energy due the spatial variation of the density 
\cite{etf,kim,manini05,sala-josephson,sala-new,recent}. 
Expanding the dispersion relation (\ref{disp}) for low momenta we get 
\beq 
\epsilon_{col}(q) = c_1 q +{\lambda \over 8 m^2 c_1} q^3  \; , 
\label{disp-low}
\eeq 
where the linear term is the familiar phonon dispersion relation 
(the so-called Bogoliubov-Goldstone-Anderson mode \cite{giorgini,levin}) 
while the cubic correction depends on both the sound velocity 
$c_1$ and the gradient parameter $\lambda$. 
Recently Escobedo, Mannarelli and Manuel \cite{cristina} 
have included an additional dispersive term in the phonon spectrum 
on the basis of the $\epsilon$ expansion of the effective 
field theory \cite{rupak2}. Here the dispersive term 
depends on $\lambda$. In general, a gradient term with $\lambda$ 
is essential to describe accurately the zero-temperature 
surface effects of a trapped system, in particular 
with a small number of atoms, 
where the Thomas-Fermi (local density, i.e. $\lambda=0$) 
approximation fails \cite{etf}. 
For the pourposes of the present paper fixing $\xi=0.42$, 
that is the Monte Carlo prediction 
for a uniform gas of Astrakharchik {\it et al.} \cite{astra}, 
we find that the best agreement with Monte Carlo 
data is obtained with $\lambda =0.25$. 

The collective modes of Eq. (\ref{disp-low}) are useful  
to describe correctly only 
the low-energy density oscillations of the system. 
At higher energies one expects the emergence of 
fermionic single-particle excitations starting 
from the threshold above which there is the breaking 
of Cooper pairs \cite{giorgini,levin,thermo-bulgac,magierski1}. 
At zero temperature these single-particle elementary excitations 
can be written as 
\beq 
\epsilon_{sp}(p) = 
\sqrt{\big({p^2\over 2m} - \zeta \epsilon_F \big)^2 + \Delta_0^2}  
\label{disp-sp}
\eeq
where $\zeta$ is a parameter which takes into account the interaction 
between fermions ($\zeta \simeq 0.9$ according to recent Monte Carlo results 
\cite{magierski1}) with $\epsilon_F$ the Fermi energy of the ideal Fermi gas.  
$\Delta_0$ is the zero-temperature gap parameter 
with $2\Delta_0$ the minimal energy to break a Cooper 
pair \cite{giorgini,levin}. The behavior of $\epsilon_{sp}(p)$ is shown 
in Fig. \ref{f1}, where we plot also (dotted line) the energy thereshold 
$\epsilon_{th}=2\Delta_0$ above which there is pair breaking and the 
continuum of single-particle excitations \cite{notarella}.  
Expanding $\epsilon_{sp}(p)$ 
around the minimum momentum $p_0=\sqrt{2m\mu}=\zeta^{1/2} p_F$, 
with $p_F=\sqrt{2m\epsilon_F}$ the Fermi momentum of the 
ideal Fermi gas, we find 
\beq 
\epsilon_{sp}(p) = \Delta_0 + {1\over 2m_0} (p-p_0)^2 \; , 
\label{sp-approx}
\eeq
where the effective mass $m_0$ is given by 
\beq 
m_0= {m\Delta_0\over 2 \zeta \epsilon_F} \; . 
\label{effective-mass}
\eeq 
Notice that the gap energy $\Delta_0$ of the unitary Fermi gas 
at zero-temperature has been calculated with Monte Carlo 
simulations \cite{carlson,magierski1} and reads 
$\gamma = {\Delta_0/\epsilon_F } \simeq 0.45$.  

\section{Elementary excitations and thermodynamics} 
  
As stressed in the introduction and in the previous section, 
at very low temperature 
the thermodynamic properties of the superfluid unitary Fermi gas can 
be obtained from the collective spectrum given by 
Eq. (\ref{disp-col}) and considering an ideal Bose gas of  
elementary excitations \cite{landau0,landau,khalatnikov}. 
As $T$ increases also the fermionic single-particle 
excitations, given by Eq. (\ref{disp-sp}) 
become important. Thus there is also the effect of 
an ideal Fermi gas of single-particle excitations. 

The Helmholtz free energy $F_0$ of the uniform ground state coincides with 
the zero-temperature internal energy $E_0$ and is given by 
\beq 
F_0 = E_0 = {3\over 5} \xi  N  \epsilon_F  \; .    
\eeq 
The free energy $F_{col}$ of the collective excitations  
is instead given by (see also \cite{landau0,landau,khalatnikov})  
\beq 
F_{col} = {1\over \beta} \sum_{\bf q} 
\ln{\left[ 1 - e^{-\beta \epsilon_{col}(q)} \right]} \; , 
\eeq
while the free energy $F_{sp}$ due to the single-particle 
excitations is 
\beq 
F_{sp} = - {2\over \beta} \sum_{\bf p} 
\ln{\left[ 1 + e^{-\beta \epsilon_{sp}(p)} 
\right]} \; .  
\eeq
Here $\beta = 1/(k_B T)$ with $T$ the absolute temperature 
and $k_B$ is the Boltzmann constant. 
The total free energy $F=F_0+F_{col}+F_{sp}$ reads 
\beq 
F = N \epsilon_F \Phi\left({T\over T_F}\right) \; , 
\label{free} 
\eeq 
where $\Phi(x)$ is a function of the scaled temperature $x=T/T_F$, 
with $T_F=\epsilon_F/k_B$, given by 
\beqa 
\Phi(x) &=& {3\over 5}\xi + {3\over 2} x \int_0^{+\infty} 
\ln{\left[ 1 - e^{-{\tilde \epsilon}_{col}(\eta)/x} \right]}
\eta^2 d\eta 
\nonumber
\\
&-& 3 x \int_0^{+\infty} 
\ln{\left[ 1 + e^{-{\tilde \epsilon}_{sp}(\eta)/x}\right]} \eta^2 d\eta 
\; . 
\label{free-scaled} 
\eeqa
Notice that the discrete summations have been replaced by integrals, 
${\tilde \epsilon}_{col}(\eta)=\sqrt{\eta^2(\lambda \eta^2+4\xi/3)}$, 
and ${\tilde \epsilon}_{sp}(\eta)=\sqrt{(\eta^2-\zeta)^2+\gamma^2}$. 
We observe that, by using the expansions 
(\ref{disp-low}) and (\ref{sp-approx}) 
for the elementary excitations, 
adopting the Maxwell-Boltzmann distribution 
for fermionic single-particles instead of the Fermi-Dirac one, 
and under the further assumption that $\lambda=0$, this formula becomes 
exactly the simple model 
\beq 
\Phi(x) \simeq {3\over 5}\xi  - 
{\pi^4 \sqrt{3} \over 80 \ \xi^{3/2}} x^4 
-{3\sqrt{2\pi}\over 2} \zeta^{1/2} \gamma^{1/2} x^{3/2} e^{-\gamma/x} \; .  
\label{free-bdm} 
\eeq
proposed Bulgac, Drut and 
Magierski \cite{thermo-bulgac}. We call this equation the BDM model.   

From the Helmholtz free energy $F$ we can immediately obtain the 
chemical potential $\mu$, that is defined as 
\beq 
\mu = \left({\partial F \over \partial N}\right)_{T,V} \; . 
\eeq
The chemical potential reads 
\beq
\mu = \epsilon_F \Big[ {5\over 3} \Phi \left({T\over T_F}\right) 
-{2\over 3} {T\over T_F} \Phi'\left({T\over T_F}\right) \Big] 
\; , 
\label{chemical}
\eeq
where $\Phi'(x)={d\Phi(x)\over dx}$ and one recovers 
$\mu_0=\xi \epsilon_F$ in the limit of zero-temperature. 

The entropy $S$ is related to the free energy $F$ by the formula
\beq 
S = - \left({\partial F\over \partial T}\right)_{N,V} \; , 
\eeq
from which we get 
\beq 
S = - N k_B \Phi'\left({T\over T_F}\right) \; .    
\label{entropy}
\eeq 
In addition, the internal energy $E$, given by 
\beq 
E = F + T S \; , 
\eeq 
can be written explicitly as 
\beq 
E = N \epsilon_F 
\left[ \Phi\left({T\over T_F}\right) - {T\over T_F} 
\Phi'\left({T\over T_F}\right) \right] \; .   
\label{internal}
\eeq

\begin{figure}
\includegraphics[width=1.\linewidth,clip]{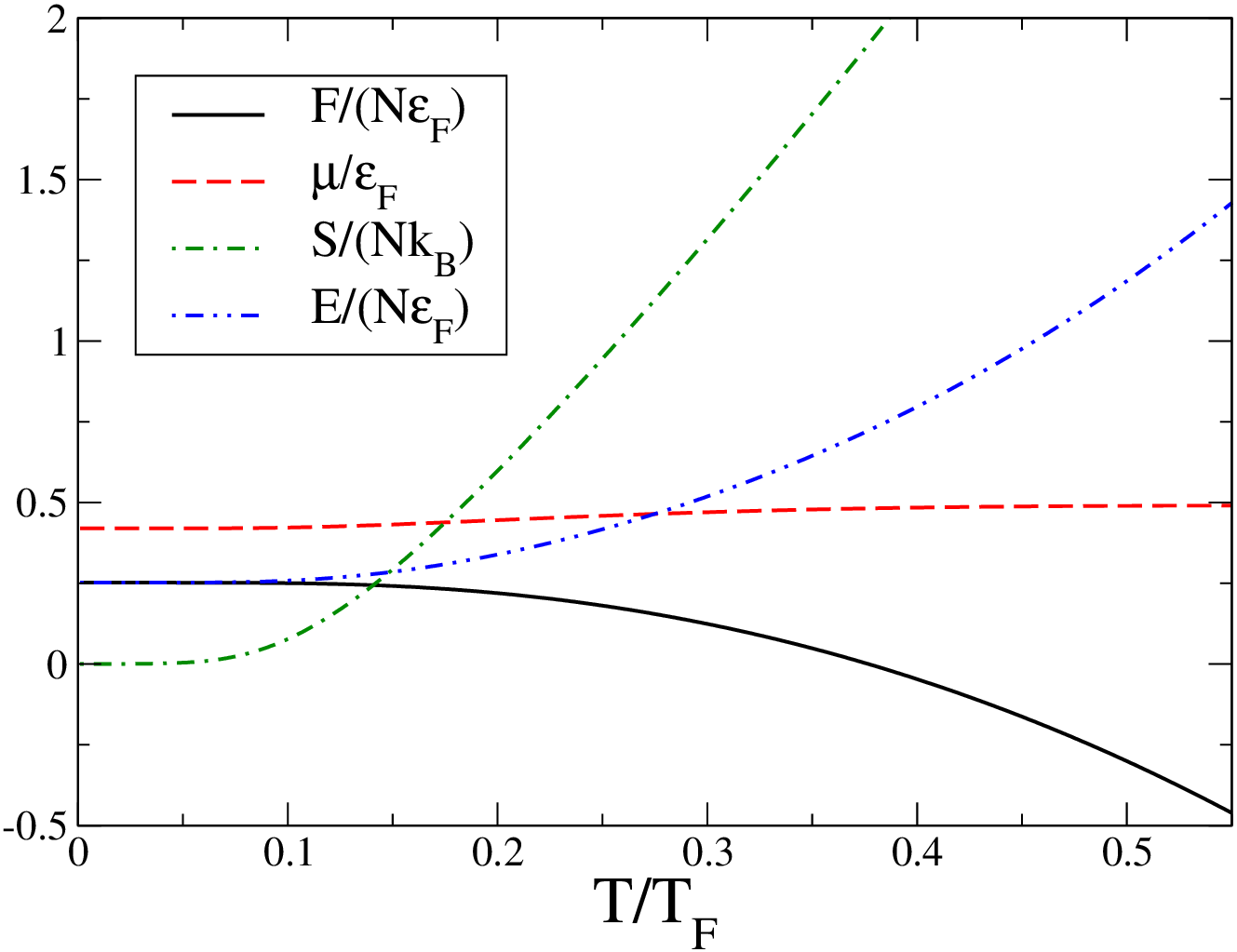}
\caption{(Color online). Thermodynamical quantities 
of the unitary Fermi gas deduced from our model. 
Zero-temperature parameters of elementary 
excitations: $\xi=0.42$, $\lambda=0.25$, $\zeta=0.9$, 
and $\gamma=0.45$.} 
\label{f2}
\end{figure}

To conclude this section we observe that the pressure $P$ of the 
unitary Fermi gas is related to the free energy $F$ by the simple 
expression 
\beq 
P = - \left( {\partial F\over \partial V} \right)_{N,T} \; . 
\eeq
We can then write the pressure as 
\beq 
P = {2\over 3} n \epsilon_F 
\left[\Phi({T\over T_F}) - {T\over T_F} 
\Phi'({T\over T_F}) \right] \; . 
\label{pressure}
\eeq
In Fig. \ref{f2} we plot various thermodynamical quantities 
obtained with our model, Eq. (\ref{free-scaled}), 
as a function of the scaled temperature $T/T_F$: 
the scaled free energy $F/(N\epsilon_F)$, the scaled 
entropy $S/(Nk_B)$, the scaled chemical 
potential $\mu/\epsilon_F$ and the scaled 
internal energy $E/(N\epsilon_F)$. 

\subsection{Gas of dilute and ultracold atoms}

It is interesting to compare our model, given by Eqs. (\ref{free}) 
and (\ref{free-scaled}), with other theoretical 
approaches and also with the available experimental data. 

\begin{figure}
\includegraphics[width=1.\linewidth,clip]{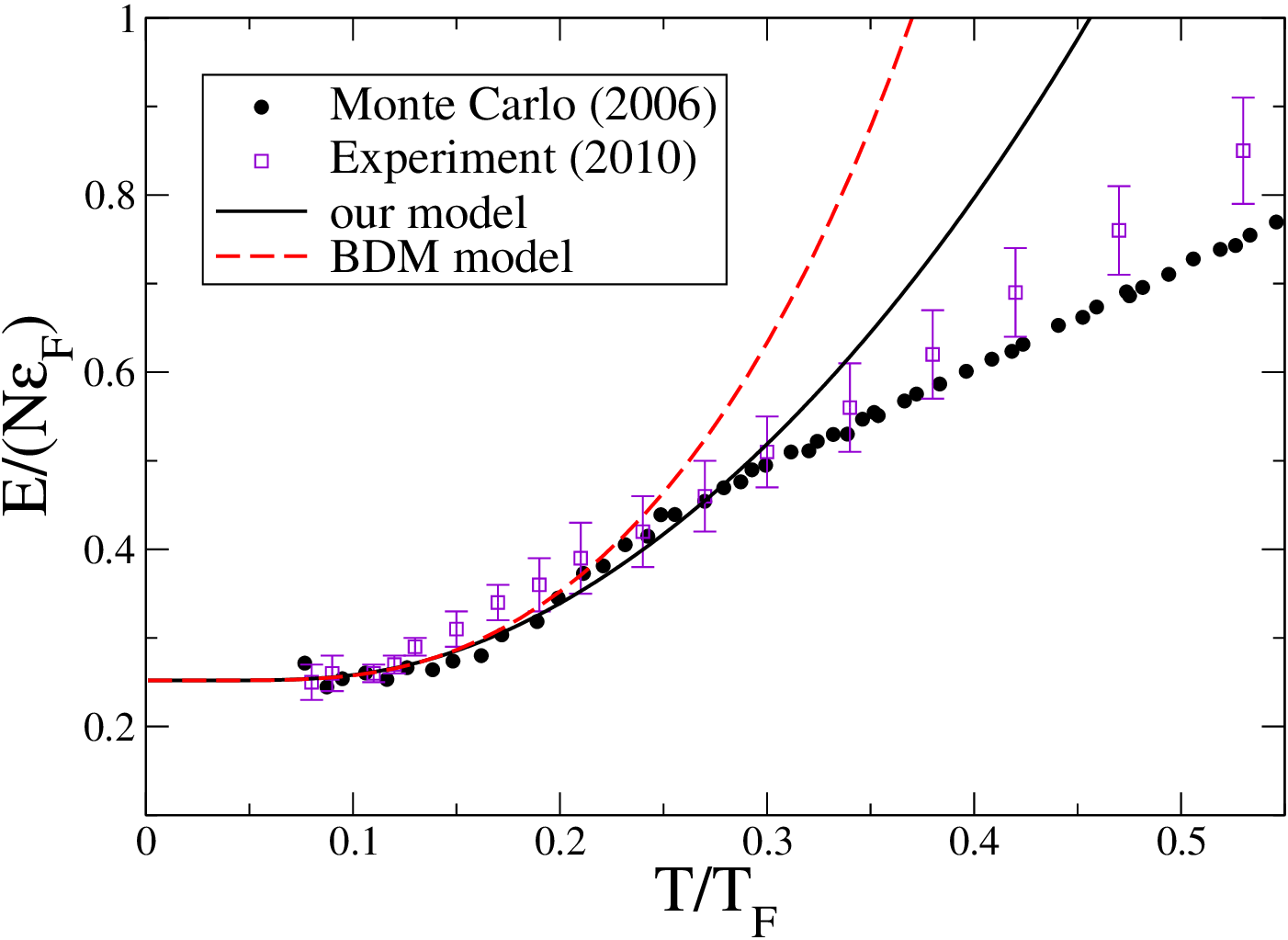}
\includegraphics[width=1.\linewidth,clip]{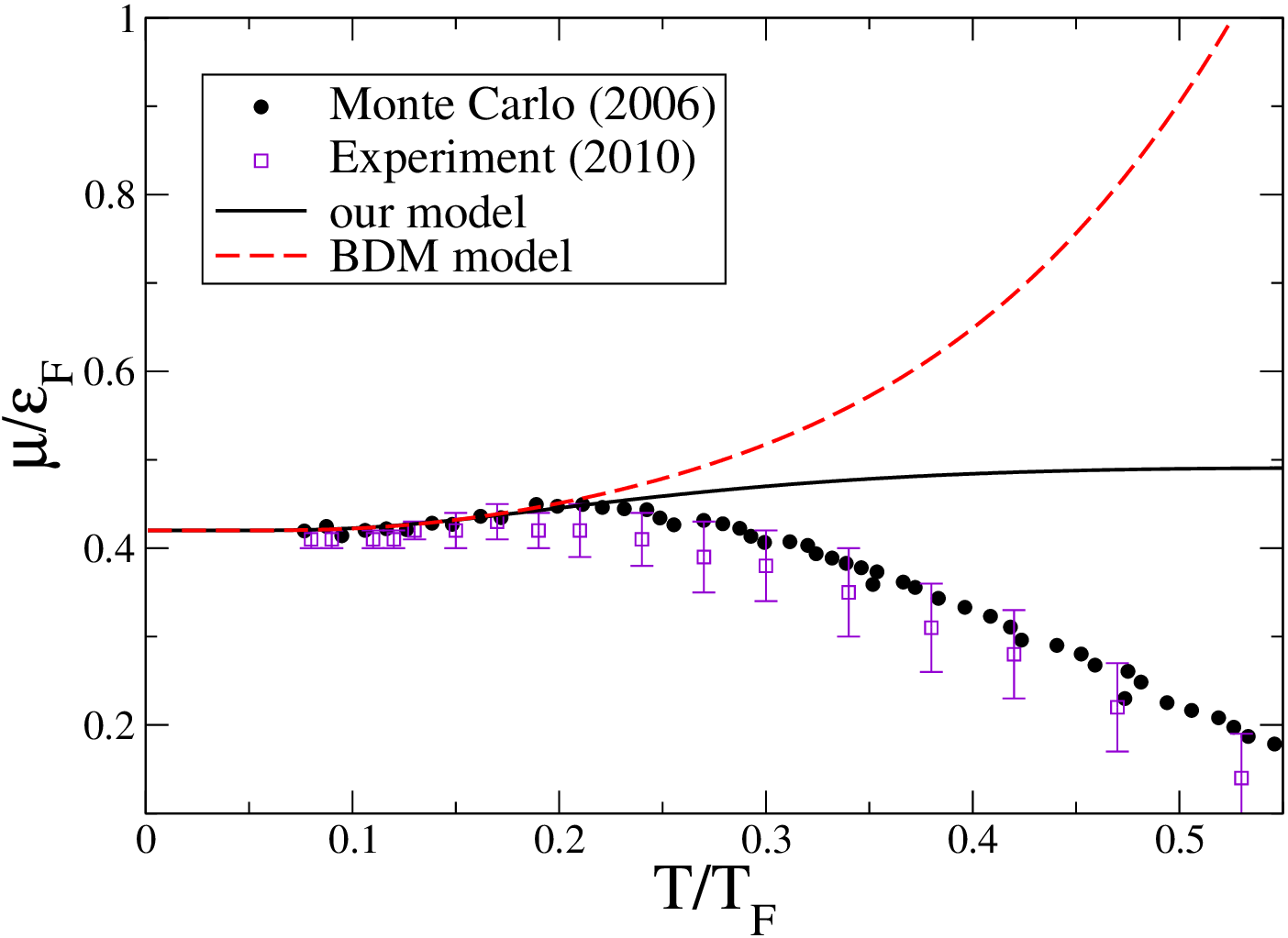}
\caption{(Color online). Atomic unitary Fermi gas. 
Upper panel: scaled internal energy $E/(N\epsilon_F)$ as a function 
of the scaled temperature $T/T_F$. 
Lower panel: scaled chemical potential $E/(N\epsilon_F)$ as 
a function of the scaled temperature $T/T_F$. 
Filled circles: Monte Carlo simulations \cite{thermo-bulgac}. 
Open squares with error bars: experimental data of 
Horikoshi {\it et al.} \cite{jap}. 
Solid line: our model, i.e. Eq. (\ref{internal}) 
with Eq. (\ref{free-scaled}). 
Dashed line: BDM model \cite{thermo-bulgac}, i.e. 
Eq. (\ref{internal}) with Eq. (\ref{free-bdm}).
Zero-temperature parameters of elementary 
excitations: $\xi=0.42$, $\lambda=0.25$, $\zeta=0.9$, and $\gamma=0.45$.} 
\label{f3}
\end{figure}

In Fig. \ref{f3} we report the data of internal energy $E$ (upper panel) 
and chemical potential $\mu$ (lower panel) obtained by Bulgac, 
Drut and Magierski \cite{thermo-bulgac} with 
their Monte Carlo simulations (filled circles) of the atomic unitary 
gas. We insert also the very recent experimental data of 
Horikoshi {\it et al.} \cite{jap} for the unitary Fermi gas of $^6$Li atoms 
but extracted from the gas under harmonic confinement (open squares 
with error bars). In the figure we include the results of 
two models: our model (solid line), that is given by 
Eqs. (\ref{internal}) and (\ref{free-scaled}); 
the BDM model (dashed line), 
that is given by Eqs. (\ref{internal}) and (\ref{free-bdm}). 

The critical temperature $T_c$ of the superfluid-normal phase transition 
has been thoretically estimated to be around $0.2 T_F$. 
In particular, the theoretical estimations are: 
$0.23$ \cite{thermo-bulgac}, $0.225$ \cite{hu}, 
$0.152$ \cite{troyer}, $0.15$ \cite{magierski1}, 
$0.245$ \cite{trivedi}, and $0.248$ \cite{wett}. 
Notice that these values are 
all much smaller than the prediction of the mean-field extended BCS 
theory which is $T_c/T_F=0.50$ \cite{giorgini,levin,hu}. 
Recent experiments with $^{40}$K  \cite{exp-new} 
and $^6$Li \cite{jap} atoms 
have measured the condensate fraction of the unitary Fermi gas 
and both suggest $T_c/T_F=0.17$. Another very recent experiment 
\cite{nascimbene} has deduced $T_c/T_F=0.157$ from the 
behaviour of the thermodynamic functions. 

Our model is based on zero-temperature elementary excitations and 
its thermodynamical quantities do not show a phase transition. Nevertheless, 
the results shown in Fig. \ref{f3} strongly suggests that our model works 
quite well in the superfluid regime, but also slightly above the 
critical temperature ($T_c\simeq 0.15$) suggested by 
two theoretical groups \cite{troyer,magierski1}. 
We have also verified that the term with $\lambda$ 
in Eq. (\ref{free-scaled}) plays a marginal role. 
The main difference between our model and the BDM model 
is instead due to the low-momentum expansions of the elementary excitations 
and to the use of the Maxwell-Boltzmann distribution instead of the 
the Fermi-Dirac one. 

\begin{figure}
\includegraphics[width=1.\linewidth,clip]{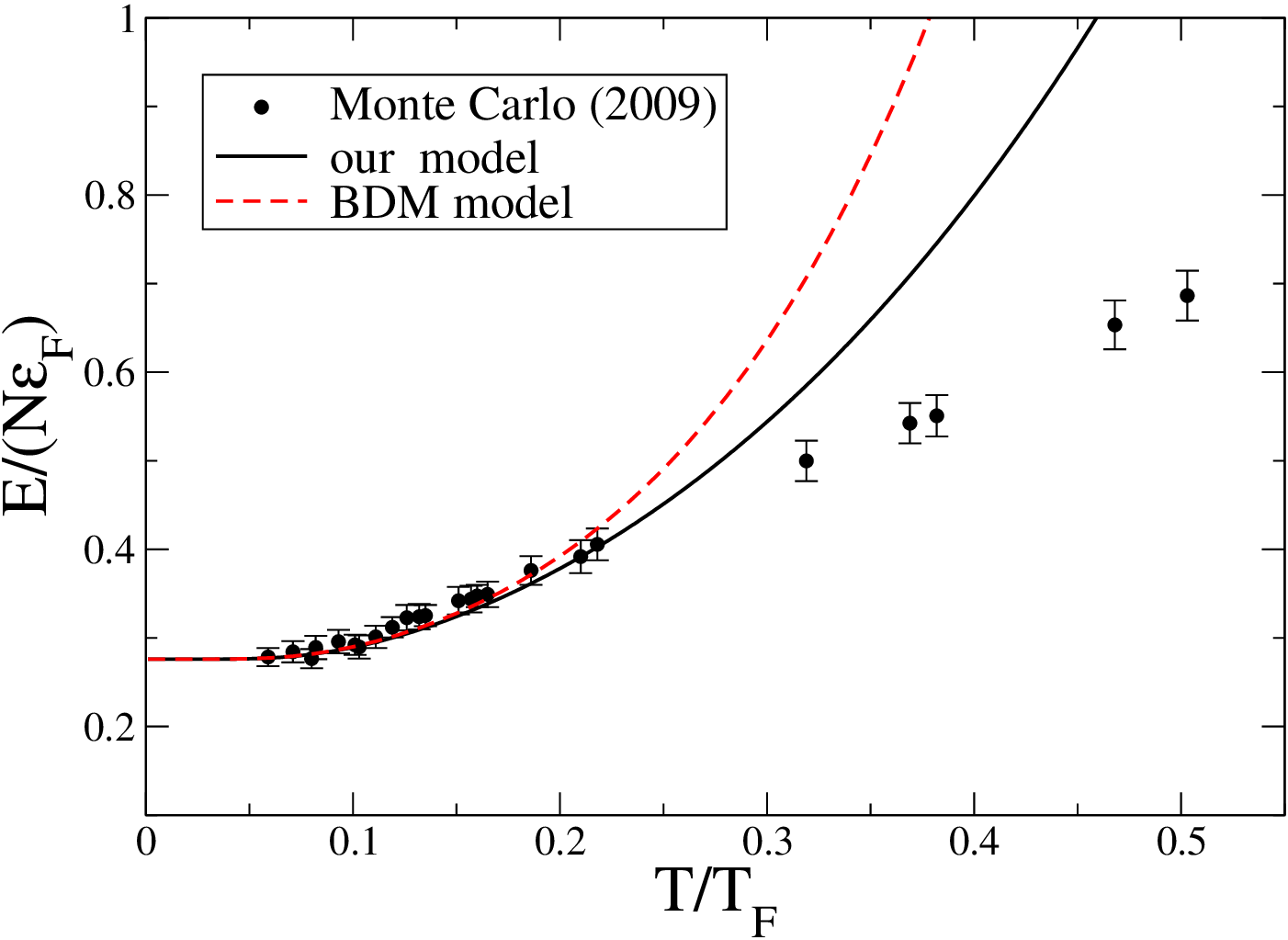}
\caption{(Color online). Dilute neutron matter at the 
density $n=0.003$ fm$^{-3}$. 
Scaled internal energy $E/(N\epsilon_F)$ as a function 
of the scaled temperature $T/T_F$. 
Filled circles: Monte Carlo simulations \cite{polish}. 
Solid line: our model, i.e. Eqs. (\ref{chemical}) and 
(\ref{internal}) with Eq. (\ref{free-scaled}). 
Dashed line: BDM model \cite{thermo-bulgac}, i.e. 
Eqs. (\ref{chemical}) and (\ref{internal}) with Eq. (\ref{free-bdm}). 
Zero-temperature parameters of elementary 
excitations: $\xi=0.46$, $\lambda=0.25$, 
$\zeta=0.82$, and $\gamma=0.29$.}  
\label{f4}
\end{figure}

\subsection{Dilute neutron matter}

Quantum Monte Carlo data of the dilute neutron matter 
close to the unitarity limit have been produced 
at finite temperature by Wlazlowski and Magierski \cite{polish}. 
The data have been obtained for the uniform neutron matter at the density 
$n=0.003$ fm$^{-3}$, where $T_F\simeq 5\cdot 10^{10}$ Kelvin 
(for comparison, $T_F\simeq 10^{-7}$ Kelvin in ultracold atomic vapors). 
In the neutron matter, the effective radius 
of the nutron-neutron interaction potential is $r_e\simeq 2.8$ fm 
and the neutron-neutron scattering length is $a \simeq -18.5$ fm. 
This means that in the calculations of Wlazlowski and Magierski \cite{polish} 
$r_e < n^{1/3}=d=6.93\ \mbox{fm} < |a|$.
Thus this dilute neutron matter is close but not equal 
\cite{pethick} to the unitarity Fermi gas ($r_e\ll d \ll |a|$). 
\cite{pethick}. Consequently, the zero-temperature parameters of the 
elementary excitations, extracted from the spectral weight function 
\cite{magierski2}, are slightly different from those of the 
unitary Fermi gas with a negligible effective range: 
$\xi \simeq 0.46$, $\zeta \simeq 0.82$, and $\gamma \simeq 0.29$ 
\cite{polish}.

In Fig. \ref{f4} we plot the scaled internal energy $E/(N\epsilon_F)$ 
versus the scaled temperature $T/T_F$ of the nuclear matter 
obtained by Wlazlowski and Magierski \cite{polish} with their 
Monte Carlo simulations (filled circles with error bars). 
On the basis of the known zero-temperature parameters of the 
elementary excitations we can compare their 
finite-temperature results with our model (solid line) 
and the BDM model (dashed line). 
This value is smaller than the one 
of the atomic unitary Fermi gas because the scaled energy gap 
$\gamma = \Delta/\epsilon_F$ of the neutron matter at $n=0.003$ fm$^{-3}$ 
is smaller than the scaled energy gap of the (atomic) unitary 
Fermi gas. Moreover, the estimated critical temperature 
for this dilute neutron matter is $T_c/T_F\simeq 0.09$ \cite{polish}. 

In agreement with the findings of Fig. \ref{f3}, 
also the results of Fig. \ref{f4} show that our model 
(solid line) works quite well in the entire superfluid regime, 
but also above $T_c$. 

\section{Superfluid fraction} 

The total number density $n$ of the unitary Fermi gas 
can be written as 
\beq 
n = n_s + n_{n} \; , 
\eeq
where $n_s$ is the superfluid density and $n_{n}$ is the normal 
density \cite{giorgini}. At zero temperature $n_{n}=0$ and $n=n_s$, 
while at finite temperature the normal density $n_n$ is finite and 
increases by increasing the temperature. Correspondingly, 
the superfluid density $n_s$ decreases and becomes equal to zero 
at a critical temperature $T_c$. The normal density is given by 
\beq 
n_{n} = n_{n,col} + n_{n,sp} \; ,  
\eeq
i.e. the sum of the normal density $n_{n,col}$ due to 
collective excitations and the normal density $n_{n,sp}$ due to 
the single-particle excitations. 
According to the Landau's approach \cite{landau0},  
the gas of collective excitations $\epsilon_{col}(p)$ which move 
with drift velocity ${\bf v}$ has a distribution 
$f_B(\epsilon_{col}(p)-{\bf p}\cdot {\bf v})$, with 
\beq 
f_B(\epsilon_{col}(p)) = {1\over e^{\beta \epsilon_{col}(p)} -1} 
\eeq
the Bose-Einstein distribution of collective excitations, 
and total linear momentum 
\beq 
{\bf P} = m \ n_{n,col}\ {\bf v} \; , 
\eeq
where the normal density $n_{n,col}$ is given by 
\cite{landau0,landau,khalatnikov} 
\beq 
n_{n,col} = - {1\over 3} \int {p^2\over m} {df_B(\epsilon_{col}(p))
\over d\epsilon_{col}} {d^3{\bf p}\over (2\pi\hbar)^3} \; . 
\eeq
Similar results hold for the normal density $n_{n,sp}$ due to 
single-particle fermionic excitations. 
It is then easy to derive the superfluid fraction 
\beq
{n_s\over n} = 1 - \Xi\left( {T\over T_F}\right) \; , 
\label{fraction}  
\eeq
where the universal function $\Xi(x)$ of the scaled 
temperature $x=T/T_F$ is given by 
\beqa 
\Xi(x) &=& {1\over x} \int_0^{+\infty} 
{e^{{\tilde \epsilon}_{col}(\eta)/x} \eta^4 \over 
(e^{{\tilde \epsilon}_{col}(\eta)/x} - 1)^2} \ d\eta \; 
\nonumber
\\
&+& {2\over x} \int_0^{+\infty} 
{e^{{\tilde \epsilon}_{sp}(\eta)/x} \eta^4 \over 
(e^{{\tilde \epsilon}_{sp}(\eta)/x} + 1)^2} \ d\eta \; ,   
\label{fraction1}
\eeqa
where ${\tilde \epsilon}_{col}(\eta)=\sqrt{\eta^2(\lambda \eta^2+4\xi/3)}$, 
and ${\tilde \epsilon}_{sp}(\eta)=\sqrt{(\eta^2-\zeta)^2+\gamma^2}$. 
The function $\Xi(x)$ can be approximated as 
\beq 
\Xi(x) \simeq  {3 \sqrt{3} \pi^4\over 40 \xi^{5/2}} x^4 + 
\sqrt{2\pi \gamma \over x} \zeta^{3/2} e^{-\gamma/x} \; ,  
\label{fraction-simple}
\eeq
by using the expansions (\ref{disp-low}) and (\ref{sp-approx}) 
for the elementary excitations, 
adopting the Maxwell-Boltzmann distribution 
for fermionic single-particles instead of the Fermi-Dirac one, 
and assuming $\lambda=0$. 

\begin{figure}
\vskip 0.5cm
\includegraphics[width=1.\linewidth,clip]{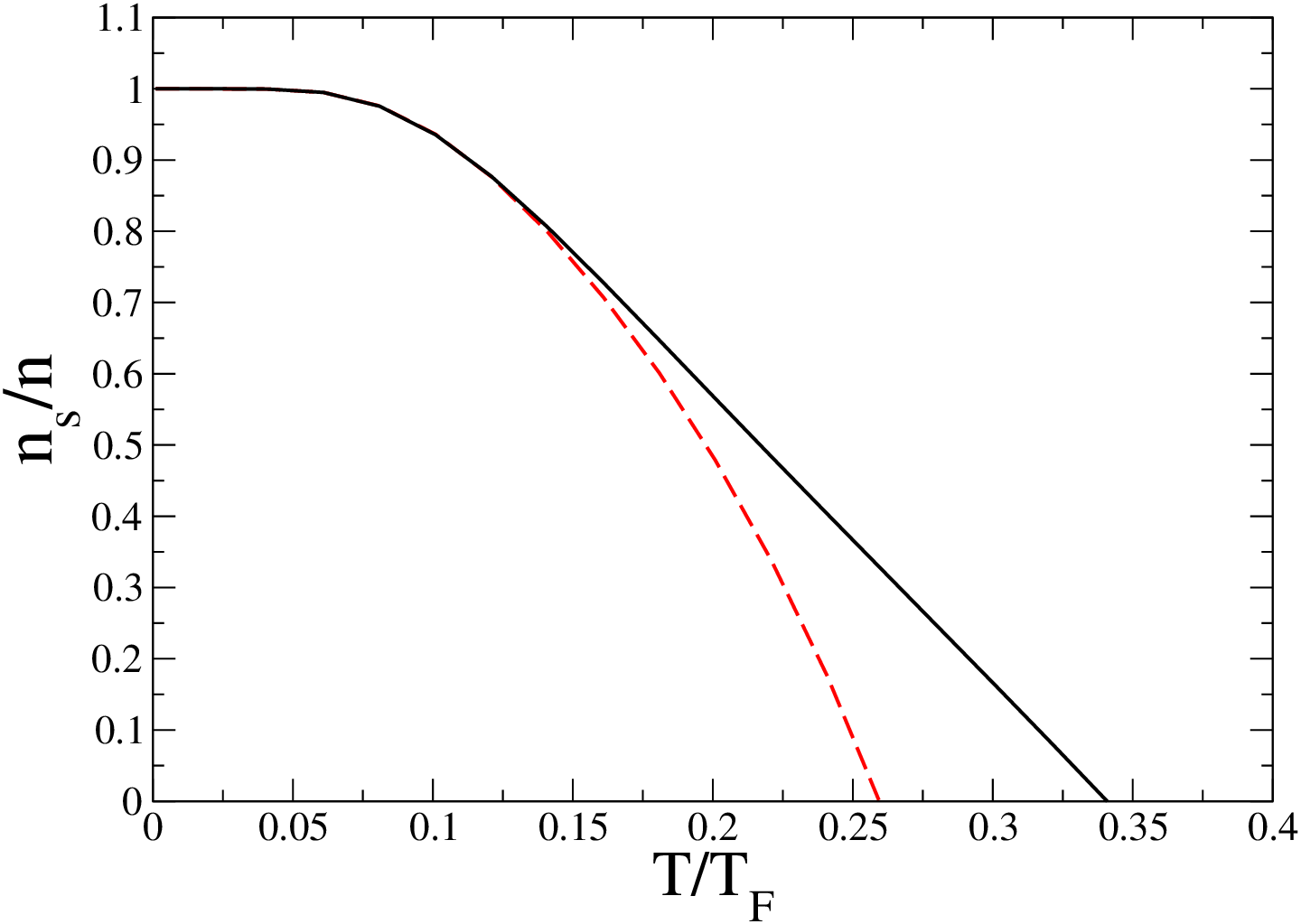}
\caption{(Color online). Superfluid fraction $n_s/n$ of the unitary Fermi gas 
as a function of the scaled temperature $T/T_F$, 
obtained by using Eq. (\ref{fraction}) with Eq. (\ref{fraction1}) 
(solid line) and Eq. (\ref{fraction-simple}) (dashed line). 
Parameters of the elementary excitations: $\xi=0.42$, $\lambda=0.25$, 
$\zeta=0.9$, and $\gamma=0.45$.} 
\label{f5}
\end{figure}

In Fig. \ref{f5} we plot the superfluid fraction $n_s/n$ of the unitary 
Fermi gas as a function of the scaled temperature $T/T_F$, 
obtained by using Eq. (\ref{fraction}) with Eq. (\ref{fraction1}) 
(solid line) and Eq. (\ref{fraction-simple}) (dashed line). 
The figure shows that 
the superfluid fraction becomes zero at $T_c/T_F= 0.34$. 
This value clearly overestimates the critical temperature with respect 
to all other beyond-mean-field determinations 
\cite{thermo-bulgac,hu,troyer,magierski1,trivedi,wett}. 
Remarkably the approximate formula, Eq. (\ref{fraction-simple}), 
is very close to the full one, Eq. (\ref{fraction}), 
up to $T/T_F\simeq 0.15$. 

\section{Sound propagation at finite temperature} 

The analysis of the sound propagation in the superfluid unitary Fermi gas 
at finite temperature can be done on the basis of the equations of 
superfluid hydrodynamics \cite{landau0,khalatnikov}, 
where superfluid and normal densities and velocities depend 
on space and time. In our problem the constitutive equations 
to be inserted in the equations of superfluid hydrodynamics 
are the Eq. (\ref{free}) of the entropy $S$ and the Eq. (\ref{pressure}) 
of the pressure $P$. 

According to Landau \cite{landau0,khalatnikov} 
any superfluid system admits a density wave, the 
first sound, where the velocities of superfluid and normal 
compoments are in-phase, and the first sound velocity is given by 
\beq 
u_1 = \sqrt{{1\over m} \left({\partial P 
\over \partial n}\right)_{\bar{S},V} } \; , 
\label{u1}
\eeq
where $\bar{S}=S/N$ is the entropy per particle. 
In addition, the superfluid system supports a temperature wave, 
called second sound \cite{landau0,khalatnikov}, where the velocities 
of superfluid and normal 
components are out-of-phase, and the second sound velocity reads 
\beq 
u_2 = \sqrt{{1\over m} {{\bar S}^2 \over 
\left({\partial {\bar S}\over \partial T}\right)_{N,V}} 
{n_s\over n_n} } \; .   
\label{u2}
\eeq 
Notice that first sound and second sound are given by 
Eqs. (\ref{u1}) and (\ref{u2}) in the hypothesis 
that these two modes are decoupled. 
As stressed by Taylor {\it et al.} \cite{taylor} 
this hypothesis is fulfilled as long as 
$R/(R+1) \ll (u_1^2-u_2^2)/(4u_1^2 u_2^2)$, 
where $R=(\bar{c}_p-\bar{c}_v)/\bar{c}_v$ is the 
Landau-Placzek ratio \cite{placzek} 
with ${\bar c}_p$ the equilibrium specific heat per unit mass  
at constant pressure and $\bar{c}_v$ the equilibrium specific 
heat per unit mass at constant density $\bar{c}_v$. 
This inequality is met also if $R$ is not small due 
to the fact that the speeds of the first and second sound of the 
unitary Fermi gas are never very close (see below). 

\begin{figure}
\includegraphics[width=1.\linewidth,clip]{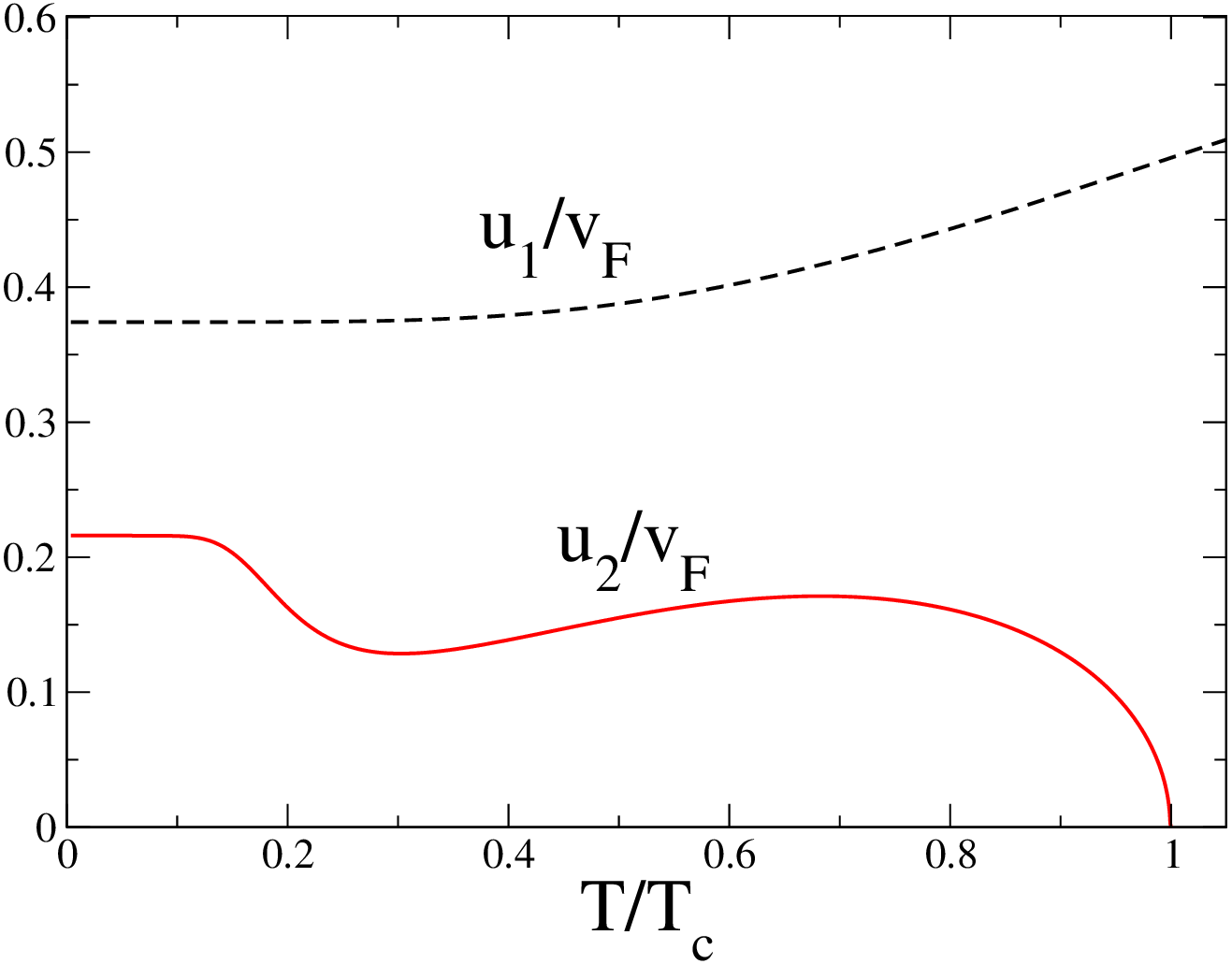}
\caption{(Color online). Dashed line: scaled first sound velocity 
$u_1/v_F$ of the unitary Fermi gas 
as a function of the scaled temperature $T/T_c$, 
obtained using Eq. (\ref{uu1}) 
with $\xi=0.42$, $\lambda=0.25$, $\gamma=0.5$, and $\zeta=0.9$. 
Solid line: scaled second sound velocity $u_2/v_F$ of the unitary 
Fermi gas as a function of the scaled temperature $T/T_c$, 
obtained using Eq. (\ref{uu2}) with $\xi=0.42$, 
$\lambda=0.25$, $\zeta=0.9$, and $\gamma=0.45$.}  
\label{f6}
\end{figure}

By using our expression (\ref{pressure}) for the pressure $P$ 
and $({\partial P\over \partial n})_{{\bar S},V} = (5/3){P/n}$ \cite{taylor} 
the finite-temperature first sound velocity becomes 
\beq 
u_1 = v_F \sqrt{ 
{5\over 9} \Phi\left({T\over T_F}\right) - 
{5\over 9} {T\over T_F} \Phi'\left({T\over T_F}\right)  } \; .   
\label{uu1}
\eeq
From this formula and Eq. (\ref{free-scaled}) it is immediate to find that 
for $T\to 0$ one has $u_1\to c_1=v_F\sqrt{\xi/3}$. 
By using our expression (\ref{entropy}) for the entropy $S$  
the finite-temperature second sound velocity can be instead written as 
\beq 
u_2 = v_F \sqrt{ -{1\over 2} {\Phi'\left({T\over T_F}\right)^2 \over 
\Phi''\left({T\over T_F}\right)} 
{1-\Xi\left({T\over T_F} \right)\over \Xi\left({T\over T_F} \right)}} \; . 
\label{uu2}
\eeq
From this formula, Eq. (\ref{free-scaled}) and Eq. (\ref{fraction}) 
with Eq. (\ref{fraction1}) 
it is not difficult to show that for $T\to 0$ one has 
$u_2\to c_1/\sqrt{3}=v_F\sqrt{\xi}/3$. 
In Fig. \ref{f6} we plot the first sound velocity 
$u_1$ and second sound velocity $u_2$ 
as a function of the scaled temperature $T/T_c$. 
These quantities are obtained 
by using Eqs. (\ref{uu1}) and (\ref{uu2}). The figure shows 
that $u_1$ is weakly dependent on the temperature $T$ 
while $u_2$ strongly depends on $T$ beteween $T=0$ 
and $T_c$, where it vanishes because $n_s=0$. These results 
are in qualitative agreement with the recent predictions 
of Taylor {\it et al.} based on a T-matrix finite-temperature 
equation of state for the unitary Fermi gas \cite{taylor}. 

\section{Conclusions} 

We have described the elementary excitations of the 
unitary Fermi gas as made of collective bosonic excitations  
and fermionic single-particle ones. This approach 
has been used many years ago by Landau with the superfluid $^4$He 
\cite{landau0} but it is also presently adopted to model other 
many-body systems, like atomic nuclei \cite{ring}. 
We stress that our approximation of non-interacting 
elementary excitations does not take 
into account the damping of collective modes, 
which becomes very important by increasing the temperature. 
We have obtained an analytical expression for 
the Helmholtz free energy and the superfluid fraction, 
showing that they are sound to study the thermodynamics of 
the unitary Fermi system, but only well below the calculated 
critical temperature of the superfluid phase transition. 
We believe that this approach to the low-temperature thermodynamics 
can be extended to the full BCS-BEC crossover of the Fermi gas 
with two equally-populated spin components. 
In this case the model requires the knowledge 
of zero-temperature elementary excitations at finite values 
of the interaction strength $1/(k_Fa_F)$. 

\noindent
{\it Acknowledgements}.  
LS thanks Flavio Toigo and Cristina Manuel for useful suggestions, 
and Aurel Bulgac, Joaquin Drut, Piotr Magierski, 
Gabriel Wlazlowski, and Munekazu Horikoshi for making 
available their data.


\begin{thebibliography}{99}
\bibitem{notarella} Actually the energy thereshold $\epsilon_{th}$ 
depends on the momentum ${\bf q}$ carried 
by the perturbation and it is given by $\epsilon_{th}(q)=2\Delta_0$ 
for $q/p_F<2\sqrt{\zeta}$ and 
$\epsilon_{th}(q)=2\sqrt{(q^2/8m-\zeta \epsilon_F)^2+\Delta_0^2}$ for 
$q/p_F>2\sqrt{\zeta}$ \cite{giorgini}.
\end{thebibliography}
\end{document}